\documentclass[12pt]{article}
\usepackage{epsfig,amsmath,graphics,float}
\usepackage{graphics,graphicx}
\usepackage{tabularx}
\usepackage{amsfonts}
\usepackage{amsmath}
\usepackage{a4}
\usepackage{color}
\def\ds{\displaystyle}

\newcommand{\beq}{\begin{equation}}
\newcommand{\eeq}{\end{equation}}
\newcommand{\lb}{\label}
\newcommand{\beqar}{\begin{eqnarray}}
\newcommand{\eeqar}{\end{eqnarray}}
\newcommand{\barr}{\begin{array}}
\newcommand{\earr}{\end{array}}

\def\XXint#1#2#3{{\setbox0=\hbox{$#1{#2#3}{\int}$}
     \vcenter{\hbox{$#2#3$}}\kern-.5\wd0}}

\def\b0{{\bf 0}}

\begin{document}

\baselineskip=18pt

\title{Experimental investigation of progressive instability and collapse of no-tension brickwork pillars\footnote{Published in 
International Journal of Solids and Structures, Volume 155, 2018, Pages 81-88, doi: 10.1016/j.ijsolstr.2018.07.010}}

\author{Massimiliano Gei\footnote{School of Engineering, Cardiff University, The Parade, Cardiff CF24 3AA, UK;
email: geim@cardiff.ac.uk.}\
               \  and Diego Misseroni\footnote{Department of Civil, Environmental and Mechanical Engineering,
University of Trento, Via Mesiano 77, I-38123 Trento, Italy; email: diego.misseroni@unitn.it.}}

\maketitle

\begin{abstract}
\noindent
The progressive instability behaviour of compressed dry-stone rectangular pillars loaded with an eccentric load is assessed experimentally and compared with the theory. Photoelastic compression tests were designed and executed on polymethyl-methacrylate brickwork pillars to reveal, i) the load-bearing capacity of the structure and the load-lateral displacement relation, ii) the effect of the eccentricity in the stress distribution along the structure, iii) the collapse mode of the system at high eccentricity. By employing a no-tension material model with linear behaviour in compression, new analytical, closed-form expressions for deformed shape of the structure, location of the neutral axis in a generic cross section and axial displacement are provided. The photoelastic stress analysis outcome fully confirms the analytical predictions for both low and high eccentricity loadings.
\end{abstract}


\section{Introduction}

The effect of loading eccentricity in masonry pillars and walls has been extensively studied experimentally in the last thirty years
\cite{hatzinikolas1,drysdale,cavaleri2005,brencich2009,adam-tony2010,sandoval2011}.
Almost all research has focused on compressed brickwork composed of bricks and mortar, a material endowed with a certain amount of tensile strength. This quantity proved to be a key property to define the load-bearing capacity of the entire structure.
%as also confirmed by finite element numerical investigations \cite{adam-tony2010,sandoval2011,brencich2005,broseghini17}.
Little is available regarding tests on dry-stone masonry pillars, for which it is sensible to assume a null tensile strength at the contact surfaces between bricks. For this type of problem, a suitable tool to predict the stability of the pillar is undoubtedly the no-tension material based on the assumption that the solid does not withstand any tensile stress \cite{delpiero_89,libro_lucchesi}. Its adoption in structural mechanics, with linear elastic behaviour in compression, was pioneered by Sahlin \cite{sahlin}, Yokel \cite{yokel}, and Frisch-Fay \cite{frisch-fay75}, that were the first to investigate instability of pillars loaded with an axial eccentric force.
Their goal was to formulate a beam-column theory for the structure accounting for the features of the constitutive model to be expressed through a
second-order differential equation having the displacement of the axis as unknown.
This problem was further studied by De Falco and Lucchesi \cite{lucchesi2002,lucchesi2003}, who also considered limited compressive strength,
and afterwards extended to circular columns by Gurel \cite{gurel16} and Broseghini et al. \cite{broseghini17}. Interestingly, both \cite{lucchesi2003,broseghini17} provided relationships, useful for design, between limit load and initial eccentricity for rectangular and circular cross sections, respectively.
Extension to nonlinear constitutive behaviour was addressed in \cite{romano93,lamendola97}, whereas
an algebraic formulation of the model based on discretisation of the pillar into a finite number of elements was proposed in \cite{lamendola93}.

%was later on paralleled to that based on a discretization of the pillar in a finite number of blocks. The latter approach requires the solution of algebraic equations formulated by imposing the equilibrium of each element \cite{lamendola93,gurel16}.

%The mechanical response of axially compressed brick masonry walls and columns is strongly affected by the eccentricity of the load. This dependency, was investigated experimentally earlier in \cite{hatzinikolas1,drysdale} and recently in \cite{cavaleri2005,brencich2008,brencich2009,adam-tony2010,sandoval2011}. As far as modelling is concerned, this problem can be tackled by assuming different hypotheses formulated to allow for the main features of the masonry whose response is strongly nonlinear: a high resistance in compression, a low tensile strength, the brittle response of the mortar, and the friction between blocks. When a continuous beam-like model is required, the assumption that the material is linearly elastic in compression and without any resistance in tension (the so called no-tension material model) can be postulated. In the past, many researchers took these hypotheses for granted to estimate failure modes and loads of many compressed slender masonry structures.

Starting from the state of the art, this paper provides three main contributions to the topic:
\begin{itemize}
\item	validate the theoretical predictions of the limit load for a prismatic brick pillar with rectangular cross section loaded with an eccentric compressive force (provided in \cite{sahlin,yokel,frisch-fay75,lucchesi2002,lucchesi2003} and based on no-tension material assumptions) performing some experimental tests on small-scale structures made up of transparent polymethyl-methacrylate (PMMA)
 bricks. A photelastic analyser was used to display the stress distribution in the pillar;
\item provide new analytical, closed-form expressions for the deformed shape of the structure, the location of the neutral axis in a generic cross section and the axial displacement for various loading eccentricities;
\item	investigate both experimentally and analytically the failure mechanisms of such pillars and the evolution of the collapse in the case of high eccentricity.
\end{itemize}

\section{Solutions based on the no-tension material model}
\lb{sect_gov_eq}

\begin{figure}[t]
	\centering
	\includegraphics[width=.8\textwidth]{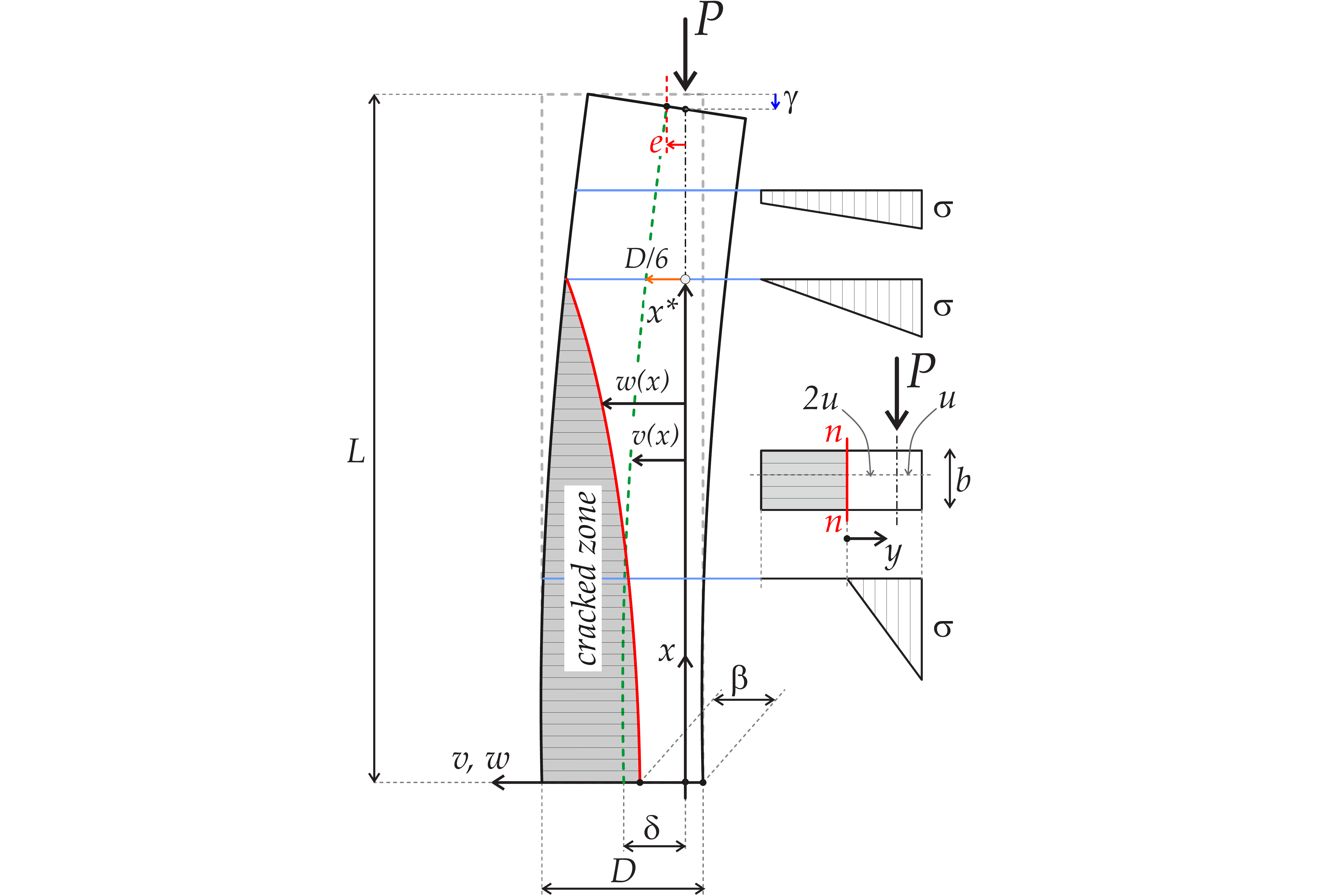}
	\caption{\footnotesize  Geometry and notation used in the analysis. The case in which the pillar is partially damaged is displayed [case iii) in the text]. The generic cross section is sketched on the right-hand side. Letter $n$ indicates the neutral axis, $\beta=3 w(0)/2$ is the height of the compressed part of the cross section at $x=0$ and $\gamma$ denotes the vertical displacement of the load $P$.}
	\label{fig1}
\end{figure}

A compressed dry-stone prismatic pillar made up of a material obeying a no-tension constitutive model may experience three regimes which depend primarily on the load eccentricity: i) the pillar is compressed everywhere; ii) each cross section of the structure is partially damaged; iii) the pillar is fully compressed in the top part along the longitudinal axis and cracked in the remaining. In particular, in our experimental study we are interested to cases ii) and iii). The equations governing these problems have been solved in closed-form for a rectangular cross section
\cite{sahlin,yokel,frisch-fay75,lucchesi2002,lucchesi2003} and with approximate methods for a circular cross section \cite{broseghini17} for a linear elastic response in compression. In Fig. \ref{fig1}, the geometry and adopted notation are introduced, in particular, $b$ and $D$ represent breadth and height of the cross section, respectively, and $x$ the coordinate taken along the axis of the pillar that coincides with the line of action of the load. Moreover, as evidenced in the same figure, the compressive stress in the reacting part of a cross section experiencing cracking takes the form
\beq
\sigma(x,y)=\frac{2}{9}\frac{Py}{bu(x)^2},
\eeq
where $y$ is the coordinate taken from the neutral axis and $u(x)$ is the distance between $x$ and the right-hand outer boundary of the cross section.

The structure is clamped at section $x=0$ and is free at the top ($x=L$), where the compressive load $P$ is applied with a given eccentricity $e$. Indicating with $v(x)$ the displacement function
of the longitudinal axis, the load-displacement curve $P(\delta)$, where $\delta=v(0)$, can be calculated for all problems listed above by imposing the following boundary and compatibility conditions
\beq
\lb{boundary_compatibility}
v'(0)= 0,\ \ \ v(L) = e,\ \ \ v(0) = \delta.
\eeq
For a partially cracked pillar, the continuity conditions
\beq
\lb{boundary_continuity}
v_-(x^*)= v_+(x^*)=D/6,\ \ \ \ v_-'(x^*)= v_+'(x^*)
\eeq
should be added at the transition coordinate $x^*$.

On the one hand, in parts of the pillar where the cross sections are fully compressed and any of these is damaged, the function $v(x)$ obeys the equation
\beq
v''(x)+\frac{P}{EJ} v(x)=0,
\lb{goveq_integro}
\eeq
where $E$ is the Young's modulus and $J = b D^3/12$ the second moment of area.

On the other, in parts where the cross sections are partially cracked, the governing equation can be represented as
\beq
v''(x)+\frac{P}{ES_n(v(x))} v(x)=0,
\lb{goveq_damage}
\eeq
where $S_n(v(x))$ is the first moment of area of the compressed portion of the cross section computed with respect to the neutral axis $n$ (Fig. \ref{fig1}).

Interestingly, the solutions in closed form to cases i) -- iii) and the determination of the expression $P(\delta)$ for each of them is facilitated by noting that $u(x)=D/2-v(x)$ and integrating once \eqref{goveq_integro} and \eqref{goveq_damage} to obtain
\beq
[v'(x)]^2=-\frac{P}{EJ} v(x)^2+c
\eeq
and
\beq
\lb{vprimop}
[v'(x)]^2=\frac{2}{27}\frac{PD^3}{EJ} \left(\frac{1}{D-2\delta}-\frac{1}{D-2v(x)}\right),
\eeq
respectively, where $c$ is an arbitrary constant; in the latter, the condition \eqref{boundary_compatibility}$_1$ has been already exploited as relevant for cases ii) and iii).

Starting from the previous equations, and by noticing that
an overbar signifies here and henceforth non-dimensionalisation with respect to the length $D$,
it is not difficult to obtain the solutions to problems i) -- iii):
\begin{itemize}
\item for case i), it turns out that the load-displacement relation $P(\delta)$ can be written in dimensionless form as
\beq
L \sqrt{\frac{P}{E J}}=\arccos (\bar e/\bar \delta);
\lb{pdeltapiena}
\eeq

\item for a pillar with all cross sections partially cracked [case ii)], the load-displacement curve can be expressed in dimensionless form as
\beq
L \sqrt{\frac{P}{E J}}=q \sqrt{1-2\, \bar{\delta}}\; T(\bar e;\bar \delta),
\lb{pdeltacracked}
\eeq
where $q=3 \sqrt{3}/(2 \sqrt{2})$ and
\beq
T(z;\bar \delta)=\sqrt{2(1-2 z)(\bar \delta-z)} +(1-2\, \bar{\delta})\,\mbox{\rm arctanh}\!
\left(\sqrt{2}\sqrt{\frac{\bar \delta-z}{1-2z}}\right).
\eeq
The deformed shape of the axis can be represented implicitly as
\beq
x(v)=L\frac{T(\bar v;\bar \delta)}{T(\bar e;\bar \delta)},
\lb{def_crack}
\eeq
and it is interesting to note that the locus of points in the plane $(x,w)$ corresponding to the \emph{neutral} axes can be detected analytically as
\beq
x(w)=L\frac{T[(1-\bar w)/2;\bar \delta]}{T(\bar e;\bar \delta)}.
\lb{neutral_crack}
\eeq
The height of the compressed part of the foundation cross section (i.e. $3 w(0)/2$) is denoted by $\beta$;

\item for case iii), it can be found that
\beq
L \sqrt{\frac{P}{E J}}=S(1/6;\bar \delta)-S(\bar e;\bar \delta)+q\sqrt{1-2\, \bar{\delta}}\; T(\bar e;\bar \delta),
\lb{pdeltamixed}
\eeq
where
%\beq
%S(z)=\mbox{\rm arctan}\!\left(\frac{z}{\sqrt{\ds{\frac{1}{12}\left(\frac{8}{9(1-2 \bar\delta)}-1\right)-z^2}}}\right).
%\eeq
\beq
S(z;\bar \delta)=\mbox{\rm arctan}\!\left\{z \left[\ds{\frac{1}{12}\left(\frac{8}{9(1-2 \bar\delta)}-1\right)-z^2}\right]^{-1/2}\right\}.
\eeq
In the cracked part of the pillar, the deformed shape of the axis can be represented implicitly as
\beq
x(v)=L\frac{q\sqrt{1-2\, \bar{\delta}}\, T(\bar v;\bar \delta)}{S(1/6;\bar \delta)-S(\bar e;\bar \delta)+q\sqrt{1-2\, \bar{\delta}}\,T(1/6;\bar \delta)},
\lb{def_mixed}
\eeq
while in the undamaged zone, it is
\beq
x(v)=L\frac{S(1/6;\bar \delta)-S(\bar v;\bar \delta)+q\sqrt{1-2\, \bar{\delta}}\,T(1/6;\bar \delta)}{S(1/6;\bar \delta)-S(\bar e;\bar \delta)+q\sqrt{1-2\, \bar{\delta}}\,T(1/6;\bar \delta)},
\lb{und_def_mixed}
\eeq
and the interface is located at
\beq
x^*=x(D/6)=L\frac{q\sqrt{1-2\, \bar{\delta}}\,T(1/6;\bar \delta)}{S(1/6;\bar \delta)-S(\bar e;\bar \delta)+q\sqrt{1-2\, \bar{\delta}}\,T(1/6;\bar \delta)}.
\lb{int_def_mixed}
\eeq
In the damaged part, the locus of neutral axes is
\beq
x(w)=L\frac{q\sqrt{1-2\, \bar{\delta}}\, T((1-\bar w)/2;\bar \delta)}{S(1/6;\bar \delta)-S(\bar e;\bar \delta)+q\sqrt{1-2\, \bar{\delta}}\,T(1/6;\bar \delta)}.
\lb{neutral_mixed}
\eeq
\end{itemize}
The transition between cases i) and iii) occurs for $\bar e=1/6$, for which (\ref{pdeltacracked}) and (\ref{pdeltamixed}) coincide.
The deformed shape of the pillar can be easily obtained by applying the translation maps $v \rightarrow v \pm D/2$ to expressions (\ref{def_crack}), (\ref{def_mixed}) and (\ref{und_def_mixed}) in the relevant domain.

Another piece of information that can be inferred from the model is the strain, say $\varepsilon_x$,
along the line of action of the force $P$ (i.e. axis $x$). As the stress herein is $\sigma_x=\sigma(x,2u(x))=4P/[9bu(x)]$,
it turns out that
\beq
\varepsilon_x=\frac{8}{9} \frac{P}{Eb \left(D-2v(x)\right)}.
\eeq
For a fully cracked pillar this quantity can be integrated along the axis $x$ to give $\gamma$ (Fig. \ref{fig1}) as
\beq
\gamma=\frac{2}{27}\frac{P D^3}{EJ}\int^{L}_0 \frac{dx}{D-2v(x)},
\label{gamma_eq}
\eeq
an equation that, coupled to eq. \eqref{vprimop}, can be solved numerically.

For a fully compressed column [case i)], the axial displacement can be easily determined by superimposing contributions from axial force and bending moment, to yield
\beq
\gamma=\frac{1}{12}\frac{P}{EJ} L D^2+\frac{1}{4}\delta^2\sqrt{\frac{P}{EJ}}\left[2L \sqrt{\frac{P}{EJ}}+\sin\left(2L \sqrt{\frac{P}{EJ}} \right) \right],
\label{gamma_fully_compr}
\eeq
where $\delta$ is functions of $e$ and $P$ through eq. \eqref{pdeltapiena}.
It is worth noting that when the foundation cross section is at the limit of cracking ($\delta=D/6$), then \eqref{gamma_fully_compr} provides
\beq
\gamma=\frac{7}{72}\frac{P}{EJ} L D^2+\frac{1}{144}\sqrt{\frac{P}{EJ}}\, D^2 \sin\left(2L \sqrt{\frac{P}{EJ}} \right).
\label{gamma_fully_compr_d6}
\eeq
In case iii), the axial displacement can be written as
\beq
\gamma=\frac{2}{27}\frac{P D^3}{EJ}\int^{x^*}_0 \frac{dx}{D-2v(x)}+\frac{7}{72}\frac{P}{EJ} (L-x^*) D^2+\frac{1}{144}\sqrt{\frac{P}{EJ}}\, D^2 \sin\left(2(L-x^*) \sqrt{\frac{P}{EJ}} \right).
\label{gamma_eq_mixed}
\eeq

As a conclusion of this section, we report the expression of the stored strain energy $W$ in a fully cracked pillar whose density $\varphi$, due to unixial stress state in the whole body, is simply
$\varphi(x,y)=\sigma(x,y)^2 /(2E)$. Therefore, the change in energy per unit axis length is
\beq
\frac{dW}{dx}=\frac{b}{2E} \int_0^{3u(x)}\sigma(x,y)^2 dy=\frac{2}{9} \frac{P^2}{Ebu(x)},
\label{rate_ww_eq}
\eeq
while its integration along axis $x$ yields
\beq
W=\frac{1}{27}\frac{P^2 D^3}{EJ}\int^{L}_0 \frac{dx}{D-2v(x)},
\label{ww_eq}
\eeq
an expression that, in light of eq. \eqref{gamma_eq}, can be rewritten as $W=P \gamma/2$ and shows the validity of the Clapeyron's theorem in this context. Extension to other cases is straightforward. An analysis of minimum principles valid for a boundary-value problem involving an elastic, no-tension material can be found in \cite{libro_lucchesi}.

\section{Photoelastic experimental tests on prismatic rectangular dry-stone pillars}

\begin{figure}[t]
	\centering
	\includegraphics[width=1\textwidth]{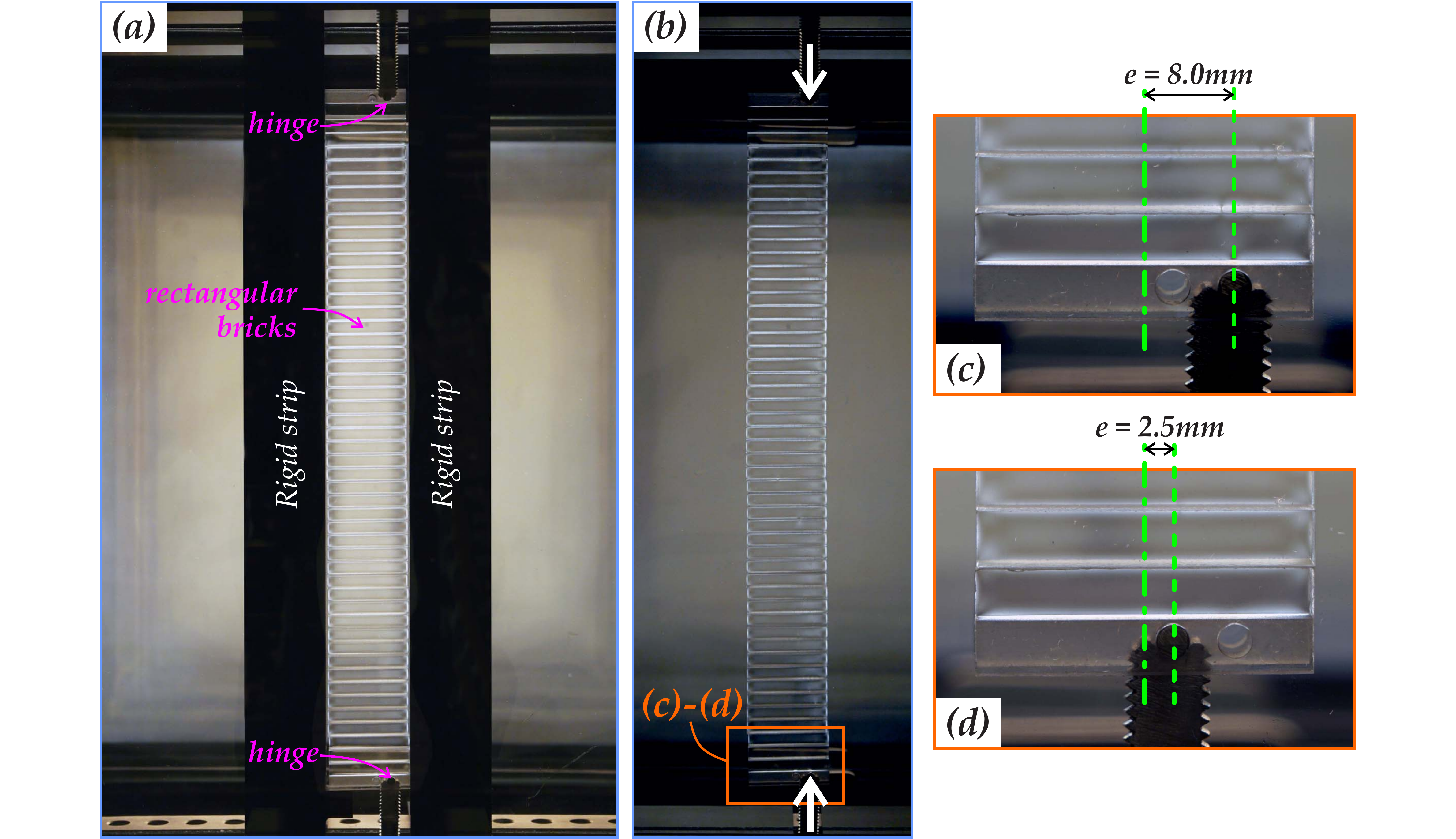}
	\caption{\footnotesize  {The experimental setup employed in the photoelastic experiments: (a) two black rigid strips are used to put bricks in position and to assure their vertical alignment, removed during execution; (b) the experimental setup at the beginning of the experiment (for an eccentricity of 8 mm); (c) and (d) details of the hinge constraint at the ends of the pillar for an eccentricity of 8 mm and 2.5 mm, respectively.}}
	\label{figsetup}
\end{figure}

Quantitative photoelastic compressive tests were performed on prismatic rectangular pillars made up of homogeneous PMMA bricks. The goals of the research were several: i) to check the predictions of load-lateral displacement equilibrium path provided by the theory, ii) to verify the actual compressive response of each cross section, iii) to look for possible partialisation by observing the internal stress distribution via photoelasticity and iv) to study the evolution of the collapse mechanism for loads with a large eccentricity.
Two different eccentricities $e$ of 8 mm  and 2.5 mm were investigated  [see Fig. \ref{figsetup}(c) and (d)]. The eccentric load was applied through two steel stinger tips connected to the testing machine on one side and with the other side duly housed and in contact with the steel cylinders (3 mm in diameter) inserted in the two ad hoc designed outer bricks of the pillar, to realize a hinged structural scheme (Fig. \ref{figsetup}). The specimen was set within two thick glasses firmly blocked on a steel frame to ensure deformation only in the bending plane. We estimated the friction between blocks and glasses to be negligible.
The bricks, of rectangular cross section ($D \times b$) of 30 mm $\times$ 10 mm (see Fig. \ref{fig1}), have been produced by drilling PMMA plates ($E=3.45$ GPa, $\rho=1200$ kg/m$^3$ and $\nu=0.35$) with an engraving machine (Roland EGX-600, accuracy 0.01 mm, manufactured by Roland).
For each eccentricity of the applied load, three different pillar configurations were considered  by changing the height of the block from 10 mm to 7 mm and eventually to 5 mm. In doing so, the tested pillars have total height $2L$ respectively of 250 mm (25 bricks), 252 mm (36 bricks) and 250 mm (50 bricks).
In total, nine complete tests were executed, two (one) for each brick height for $e=8$ mm ($e=2.5$ mm).
Afterwards, with the purpose to record the collapse mechanism at high eccentricity ($e=8$ mm), three tests, one for each height of the block (5 mm, 7 mm and 10 mm), were executed.

The load to the structure was applied by imposing a displacement of 4 $\mu$m/s on dry-stone pillars with an ELE Tritest 50 machine (ELE International Ltd) installed inside a linear and a circular (with quarter-wave plates for 560 nm) polariscope equipped with a white and sodium vapor light (Tiedemann \& Betz).  The polariscope was designed and constructed at the University of Trento \cite{giovanni2011,giovanni2011_2} and adopted in \cite{stiffener2008,noselli_fracture2010,summer2014}. In all tests, a preload of 50 N was applied to prevent the lateral pull out at the beginning of each test and to increase the friction between blocks. The force data was measured by a TH-KN2D load cell (RC 20 kN, from Gefran) directly connected to the top stinger tip and recorded by an acquisition system NI CRio interfaced to a PC by software Labview, ver. 2016 (National Instruments). The load cell was calibrated just before the experiments.
Along the execution of the first nine tests, pictures were taken by a Sony Alpha 6300 camera to derive the stress distribution and the lateral displacement of the mid cross section of the pillar. At the same time, movies were taken with a Sony PXW-FS5 video camera (24 fps). Using the same device, high speed movies (240 fps) were recorded
in order to visualise the collapse mode of the pillar loaded at high eccentricity (see Fig. \ref{collapse}).
All the experiments were performed in the Instabilities Lab at the University of Trento.

\section{Analysis of results and comparison with the theory}

\begin{figure}[t]
\centering
\includegraphics[width=1\textwidth]{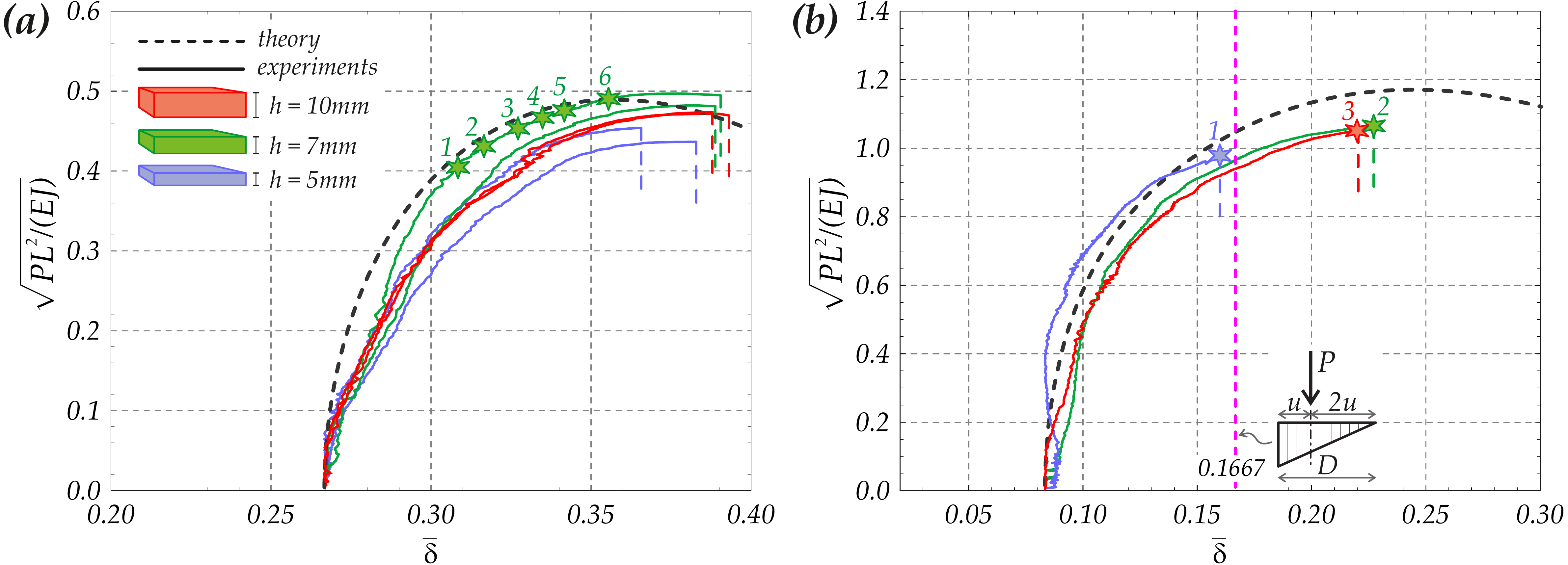}
\caption{\footnotesize
{The dimensionless axial load $L \sqrt{P/EJ}$ measured as a function of the transverse displacement $\bar \delta$ for bricks having height of 5 mm (blue), 7 mm (green), 10 mm (red) and comparison with the theoretical predictions (dashed/black line). The subfigure (a) is for an eccentricity of 8 mm [dashed line represents eq. (\ref{pdeltacracked}), markers refer to the configurations reported in Fig. \ref{sovrapp}] while subfigure (b) is for an eccentricity of 2.5 mm [dashed line represents eq. (\ref{pdeltapiena}) for $\bar \delta<1/6$, eq. (\ref{pdeltamixed}) for $\bar \delta>1/6$].}}
\label{figrissper}
\end{figure}

The dimensionless load-lateral displacement curves obtained after postprocessing of the experimental data are reported in Fig.
\ref{figrissper} with solid lines, while theoretical predictions are sketched with black/dashed lines.
In order to retrieve the data from the experiments, the recorded frame sequences described in the previous section were elaborated by a Mathematica software (ver. 11.2, Wolfram Research, Inc., Champaign, IL, 2017), which is specifically programmed for this type of experiments. This code is able to capture the deformed shape of the compressed pillar and also the horizontal displacement $\delta$ by individuating the pixel contrast between the coloured pillar and the  black background.

\begin{figure}[t!]
	\centering
	\includegraphics[width=0.9\textwidth]{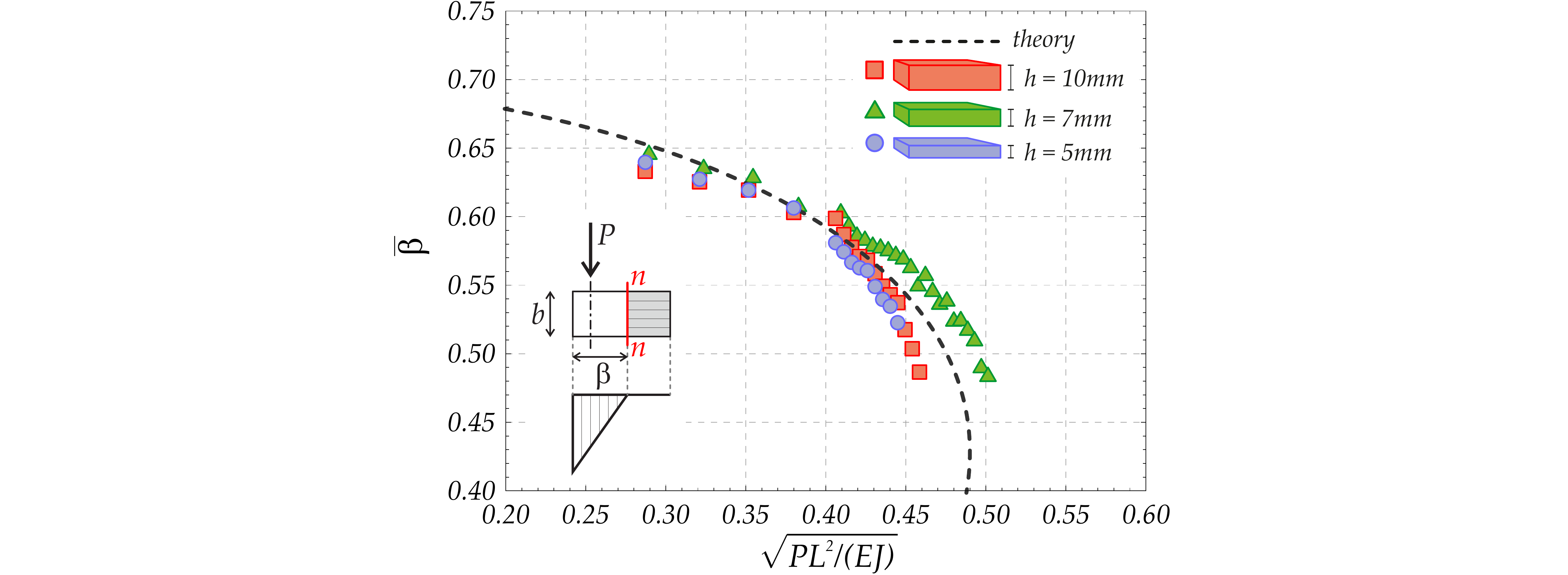}
	\caption{\footnotesize
		Experimental measure of the height of the cross section subjected to compression ($\bar \beta=\beta/D$) at mid height of pillars for an eccentricity of 8 mm and its comparison with the theoretical prediction  based on eq. (\ref{neutral_crack}). }
	\label{neutro}
\end{figure}

\begin{figure}[t]
	\centering
	\includegraphics[width=1\textwidth]{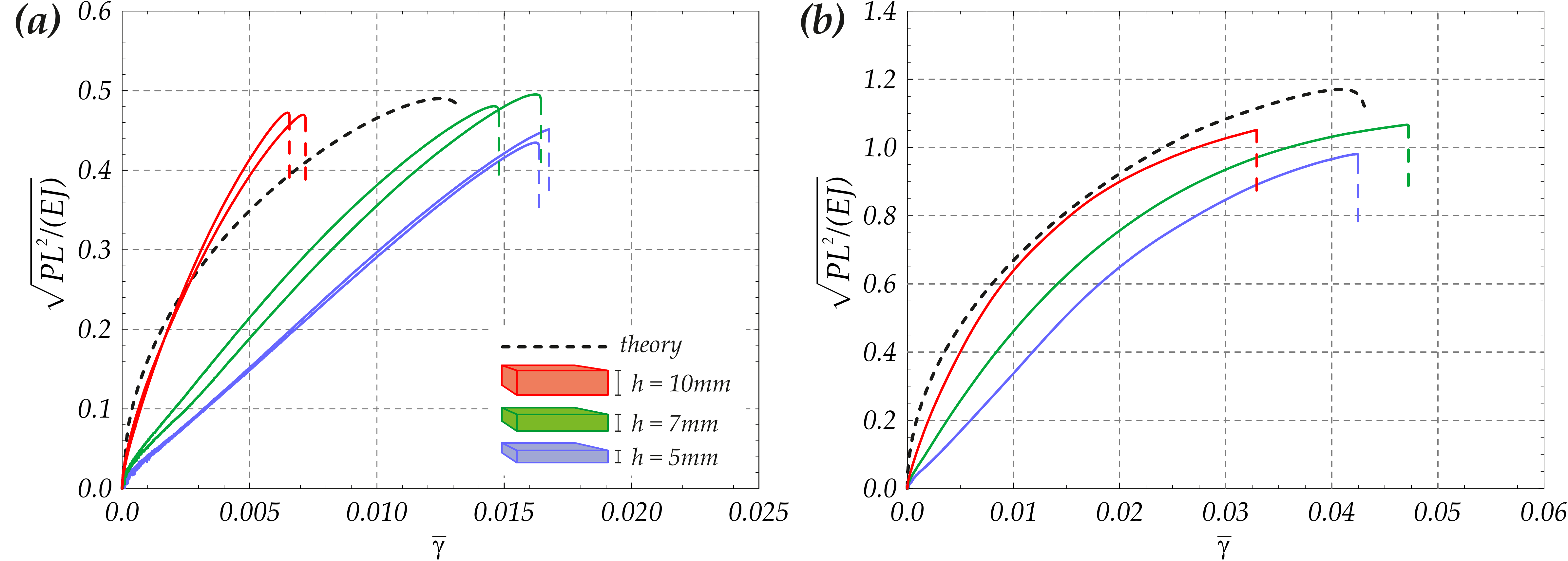}
	\caption{\footnotesize
		{Dimensionless axial load $L \sqrt{P/EJ}$ vs dimensionless vertical displacement $\bar \gamma$ at failure for bricks having height of 5 mm (blue), 7 mm (green), 10 mm (red) and comparison with the theoretical predictions (black/dashed line). (a) $e=8$ mm [the dashed line represents eq. (\ref{gamma_eq})]; (b) $e=2.5$ mm [dashed line represents eq. \eqref{gamma_fully_compr} for $\bar \delta<1/6$, eq. \eqref{gamma_eq_mixed} for $\bar \delta>1/6$].}}
	\label{figrissperaxial}
\end{figure}

\begin{figure}[t]
	\centering
	\includegraphics[width=0.85\textwidth]{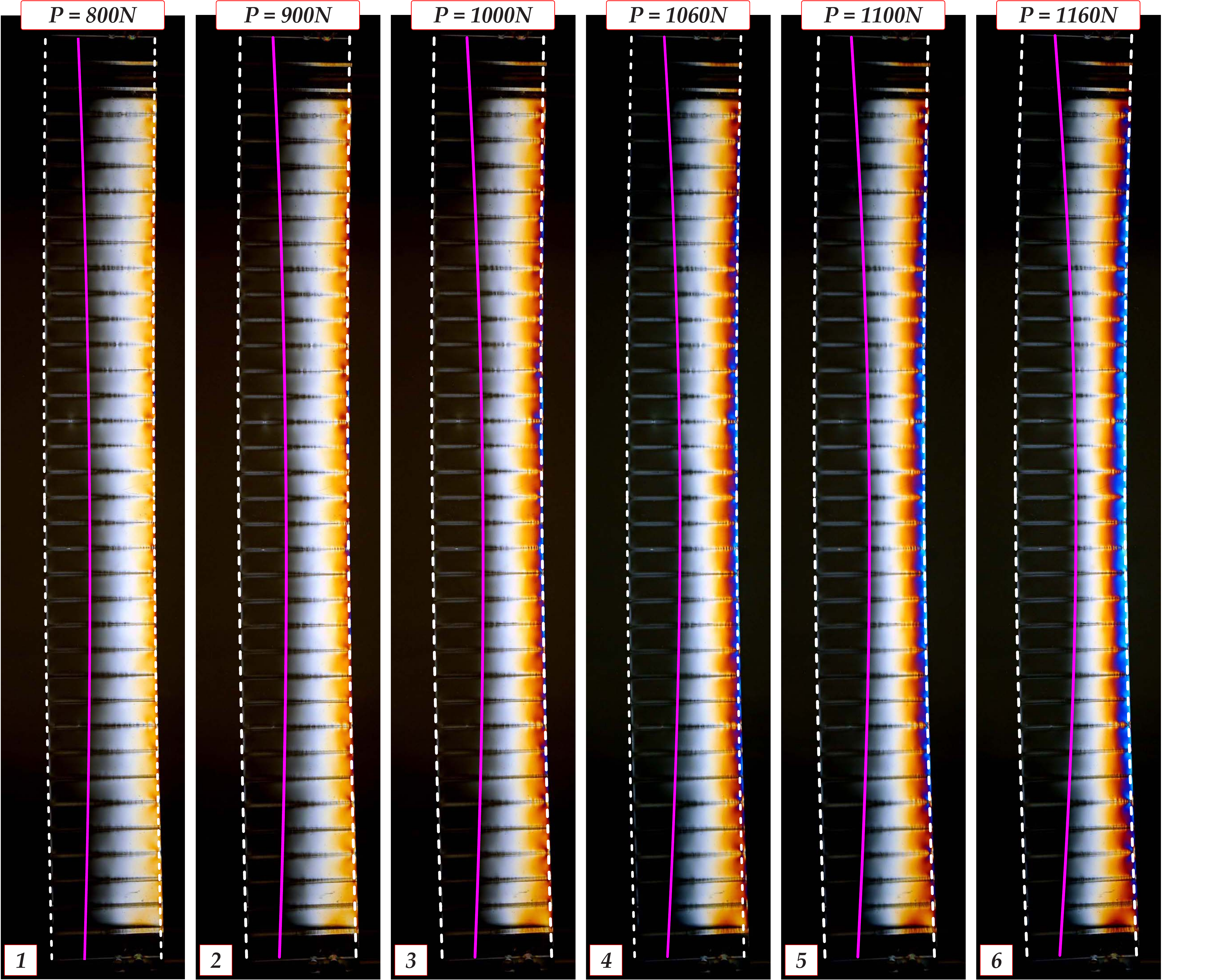}
	\caption{\footnotesize
		Stress distribution revealed by photoelastic investigations for a pillar made up of 36, 7 mm height bricks and loaded with an eccentricity of 8 mm. The coloured part in each picture corresponds to the stressed part of the pillar. Locations of neutral axes predicted by the theory, eq. (\ref{neutral_crack}), are sketched with magenta/continuous lines. Starting from the left, loads of the pictures correspond, respectively, to $L \sqrt{P/EJ}= $ 0.406, 0.431, 0.454, 0.467, 0.476, 0.489 [see markers in Fig. \ref{figrissper}(a); the same figure also shows to which experiment the pictures are referred to]. The white/dashed lines represent the deformed shape of the pillar as predicted by the theory (see Section \ref{sect_gov_eq}).}
	\label{sovrapp}
\end{figure}

Fig. \ref{figrissper}(a) refers to loads applied at the highest eccentricity ($\bar e=0.2667$), so that all cross sections of the pillar are partially cracked [case ii)]. All tests represented here reached a structural collapse mode similar to that that will be shown in Fig. \ref{collapse}. The six markers on the top experimental curve refer to configurations that will be analysed in Fig. \ref{sovrapp}. With reference to the theoretical predictions, eq. (\ref{pdeltacracked}), we note that, on the one hand, the no-tension model slightly overestimates the actual behaviour of the three pillars; on the other, the height of the brick seems not a key factor in setting the bearing properties of the structure as the pair of curves for $h=10$ mm lays between the other two pairs.
In all tests here reported, a slight adjustment of the bricks has taken place about the dimensionless load $L \sqrt{P/EJ}=0.1$.

\begin{figure}[t]
	\centering
	\includegraphics[width=0.85\textwidth]{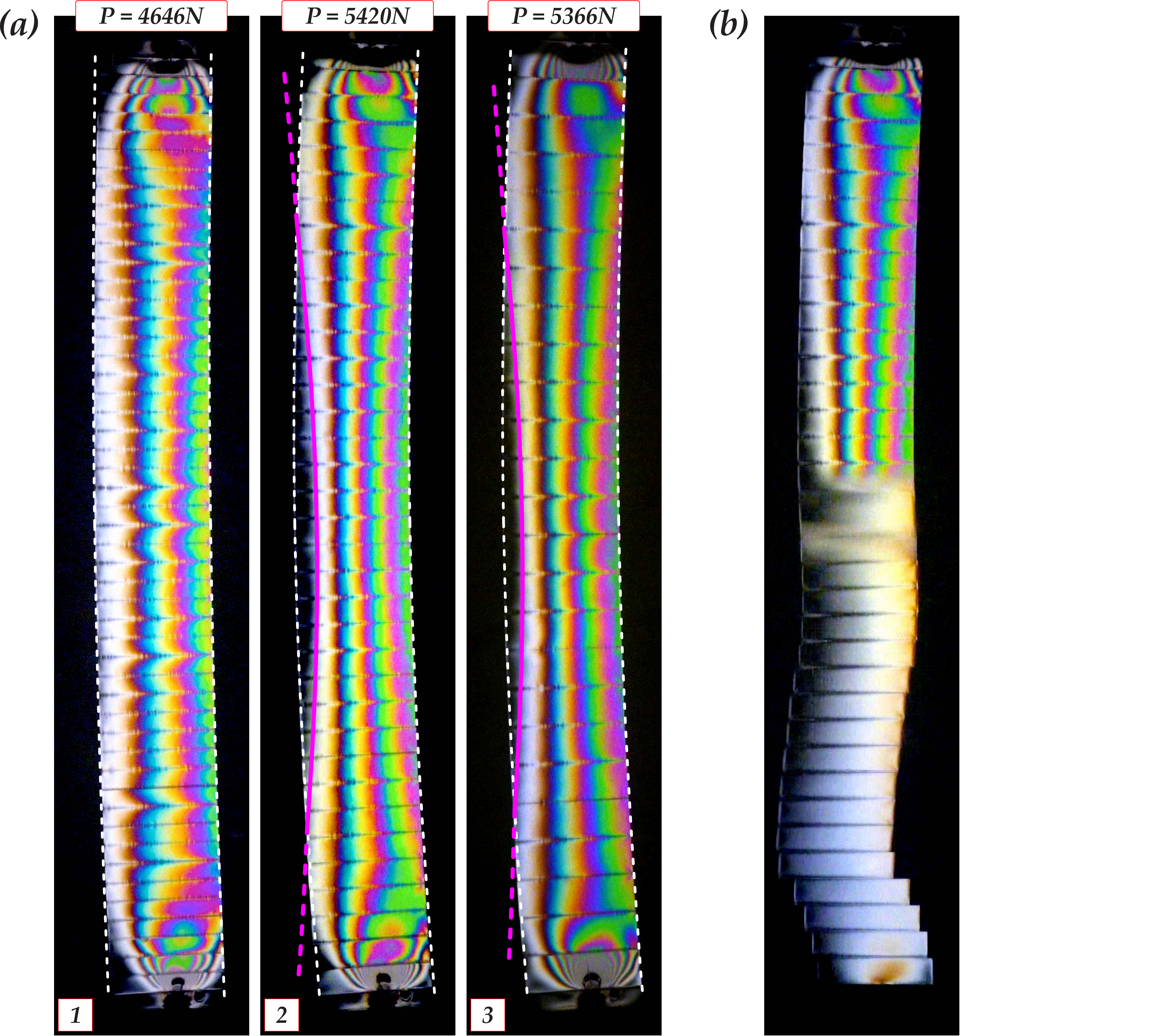}
	\caption{\footnotesize
		Stress distribution revealed by photoelastic investigations for pillars loaded with an eccentricity of 2.5 mm. The coloured part in each picture corresponds to the stressed part of the pillar. (a) Locations of neutral axes in the central part of the pillar predicted by the theory, eq. (\ref{neutral_mixed}), are sketched with magenta/continuous lines. Starting from the left, the pictures correspond, respectively, to $h=5,\ 7,\ 10$ mm, with loads given by $L \sqrt{P/EJ}= $ 0.978, 1.065, 1.051 [see markers in Fig. \ref{figrissper}(b)]. The white/dashed lines represent the deformed shape of the pillar as predicted by the theory (see Section \ref{sect_gov_eq}). (b) Picture taken just after the failure of the bottom outer brick.}
	\label{sovrapp25}
\end{figure}

Fig. \ref{figrissper}(b) is concerned with the lowest eccentricity ($\bar e=0.0833$). Load is applied within the middle-third of the outer cross sections and the pillars are fully compressed at the initial stage of the tests. All tests terminated with a local failure of the bottom outer brick as a result of the large stress concentrated
in the small contact area between the steel tip and its housing.
The three markers correspond to configurations reported in Fig. \ref{sovrapp25}.
The theoretical black/dashed line represents eq. (\ref{pdeltapiena}) for $\bar \delta<1/6$ and eq. (\ref{pdeltamixed}) for $\bar \delta>1/6$.
The line at $\bar \delta=0.1667$ marks the theoretical transition between a fully compressed pillar and the onset of cross-section partialisation.
For this set of tests, the experiments are closer to the model with respect to those displayed in subfigure (a).

%The experimental results and the theoretical prediction are in both cases ($e=8$ mm and $e=2.5$ mm) in high agreement.

Fig. \ref{neutro} gives an account of the comparison between theory [eq. (\ref{neutral_crack}) in black/dashed lines]
and experiments (disk, triangle and square markers) of the height $\beta$ of the compressed portion of the mid cross section of the pillar for high-eccentricity loading conditions. Data from tests for all the three brick heights were considered.
The agreement between the two approaches is on average very good.

An additional comparison between theory and experiments can be performed for the axial displacement $\gamma$. In Fig. \ref{figrissperaxial}, the outcome of the theoretical model (black/dashed lines) is reported together with values extracted
from experiments at failure (coloured lines). In particular, in order to discount the effect of the high strain field concentrated in the neighbourhood of the housings of the stingers, these have been calculated measuring, on the available high resolution pictures, the relative displacement between the interfaces just adjacent the two outer bricks. For the high-eccentricity case [subfigure (a)], the data are quite scattered, but the average is in good agreement with the theory. For the low eccentricity case, the model slightly underestimates the axial displacement.

The distribution of the stress in a high-eccentricity loaded pillar --and for $h=7$ mm-- revealed by the photoelastic apparatus is reported in Fig. \ref{sovrapp}. The pictures clearly highlight the presence of a non-reacting part of the pillar, the high stress localised on the right-hand edge and the deformed configuration of the sample.
The reason for which the stress is not uniform across the interface between two bricks is due to the residual stresses that the machining process invariably introduces on the surfaces of each PMMA component.
The location of the neutral axes and the deformed shapes of the pillar provided by the no-tension model are reported in each snapshot
with magenta [eq. (\ref{neutral_crack})] and white/dashed lines, respectively. The agreement is in general excellent, however we note that, by approaching the two extremes of the pillars, magenta lines slightly diverge from the apparent position of the neutral axes as the loads are applied through concentrated forces and diffuse in the specimen in the bricks adjacent to the outer ones.

Fig. \ref{sovrapp25} deals with low-eccentricity loadings. The loads of the three pictures in subfigure (a), one for each of the brick heights, correspond to those of the markers reported in Fig. \ref{figrissper}(b). In comparison with the previous figure, here the stresses are much higher as revealed by the intensity of the colour fringes. The pillar in picture no. 1 is fully compressed, as predicted by the theory, while the other two display an inner part where the cross sections are partially unstressed. The corresponding location of the neutral axes is also here pinpointed by the magenta solid lines whose analytical expression is now eq. (\ref{neutral_mixed}), whereas white/dashed curves detect the current configurations envisaged by the theory.
Subfigure (b) is a snapshot taken at failure of the pillar with $h=7$ mm. The breakage of the bottom outer brick induces a decompression wave that propagates upward along the structure.

\section{Collapse mode}

\begin{figure}[t]
	\centering
	\includegraphics[width=.95\textwidth]{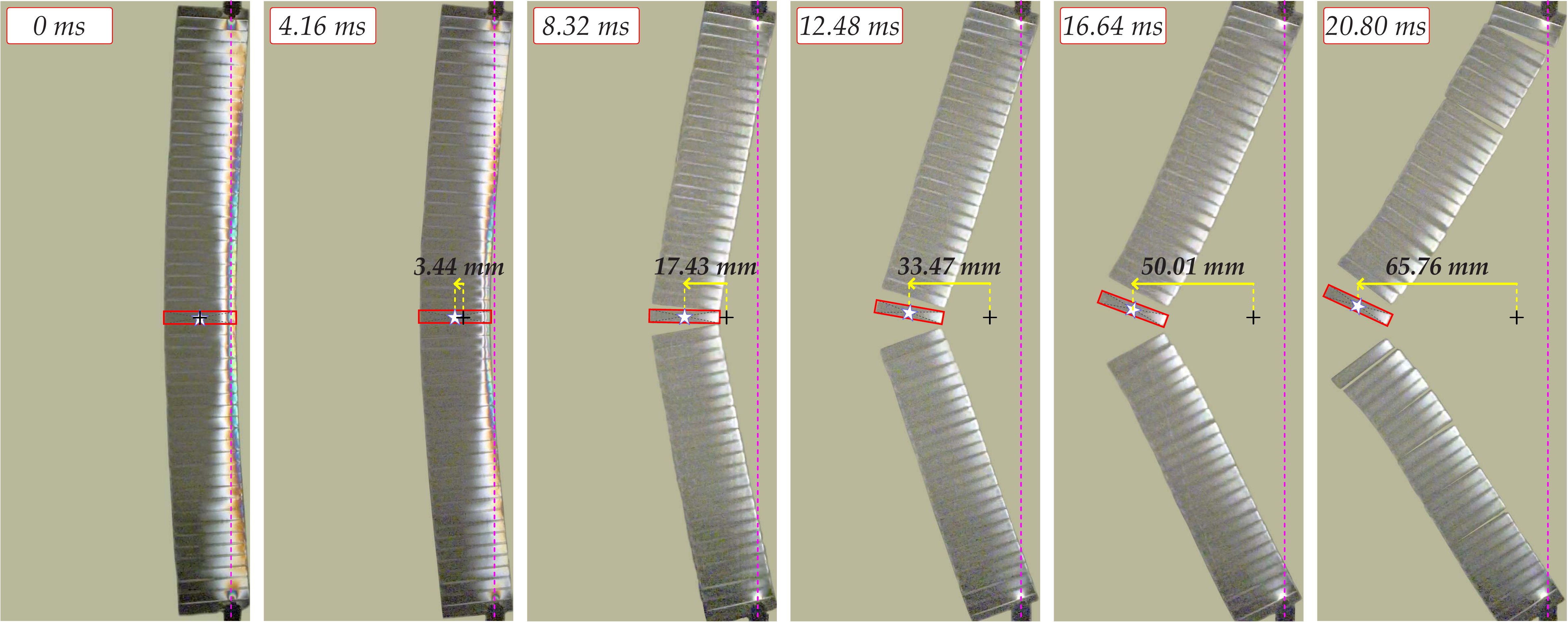}
	\caption{\footnotesize
		A sequence of snapshots (taken from a high-speed movie, 240 fps) showing the collapse mode of masonry pillar for the case of an eccentricity of 8 mm and bricks 5 mm in height. The collapse load is $P=$ 870 N. In each snapshot, the position of the centre of mass of the central brick with respect to the frame at 0 ms is reported.}
	\label{collapse}
\end{figure}

The collapse mode of one of the pillar with bricks 5 mm in height, loaded at high eccentricity, is shown in Fig. \ref{collapse}.
In the sequence, the pre-collapse state is represented in the left-hand frame at 0 ms, subjected to the load $P=870$ N. By examining in detail this picture, we have estimated that the maximum stress and the height $3u$ of the compressed part of the mid cross section are approximately $24.7$ MPa and $7.05$ mm, respectively. The collapse is associated with the loss of equilibrium of the structure and is not due to the crushing of any brick of the pillar. At this stage, the mid cross section acts as a hinge.
In the second picture from the left (4.16 ms), equilibrium of the pillar is lost as the action line of the load falls outside its contour. Therefore, collapse has just started. The picture at 8.32 ms sets the moment when the central brick loses contact with the two adjacent elements. From this point on, it approximately moves at a constant average speed of 3.87 m/s (this value can be computed by using the distances of the centre of mass indicated in the figure).
This speed, denoted by $v_b$, can be estimated by imposing conservation of energy in at least two different ways.

\begin{figure}[t]
	\centering
	\includegraphics[width=.8\textwidth]{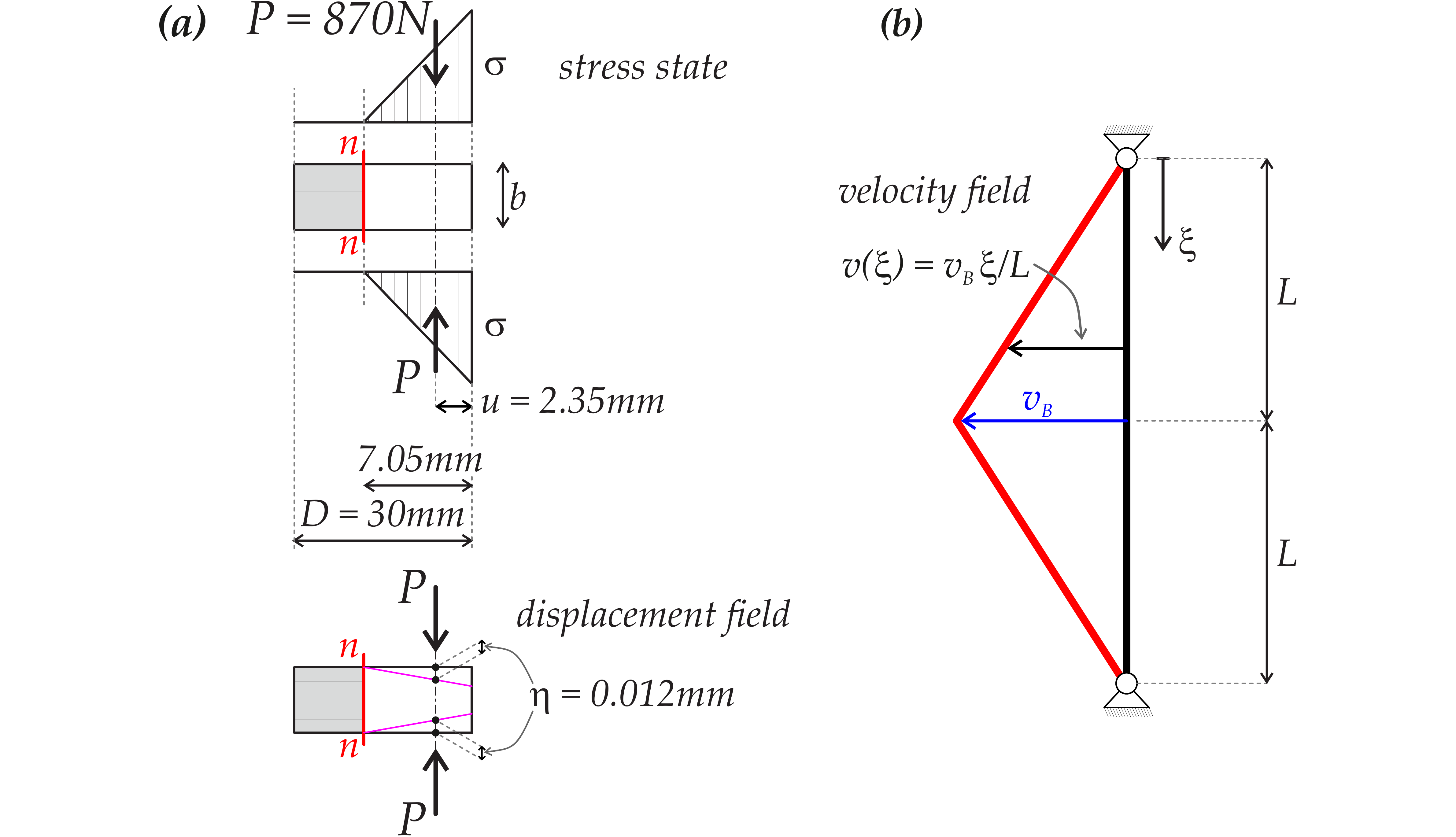}
	\caption{\footnotesize Schematics for methods to estimate the speed $v_b$ of the central brick after collapse of a pillar loaded with a force with high eccentricity. (a) Stress state and displacement field in the central brick on the verge of the collapse; (b) three-hinge type velocity distribution for the estimate of the kinetic energy in the second method.}
	\label{collapse_2}
\end{figure}

{\bf First method: analysis of the central brick.} One possibility [Fig. \ref{collapse_2}(a)] is to equate the strain energy ($W_b$) stored in the brick in the static configuration on the verge of the collapse (frame at 0 ms) to the kinetic energy ($K_b=m_b v_b^2/2$) owned by the brick itself after the structure is completely disarranged (the mass of the brick $m_b$ is equal to $1.763$ g). The quantity $W_b$ can be evaluated in two ways:
a) through eq. \eqref{rate_ww_eq}, in which case
\beq
W_b=2 P^2h/(9Ebu),
\lb{wbrick}
\eeq
and b) by means of the Clapeyron's theorem applied to the single brick, through the expression $W_b=P\eta$, where $\eta$ is half of the shortening of the part of axis $x$ belonging to the brick and estimated as $\eta=0.012$ mm from the data obtained from the picture. Both yield $W_b=10.44\cdot 10^{-3}$ J and the match between energies leads to $v_b \approx 3.44$ m/s.

{\bf Second method: analysis of the whole pillar.} The second method that is proposed assumes that the collapse is modelled through a typical structural \lq three-hinge' mechanism [Fig. \ref{collapse_2}(b)], where each half of the structure displays a linear velocity (and displacement) field $v(\xi)$ that vanishes at the two outer hinges and whose expression is $v(\xi)=v_b\, \xi/L\ (0\leq\xi\leq L)$. The total strain energy ($W_p$) in the pillar pictured in the frame at 0 ms is then converted to the kinetic energy $K_p$ associated with the mechanism.
The former can be estimated summing up all contributions of the type \eqref{wbrick}, namely $W_p=2 P^2h/(9Eb)\sum_{k=1}^{48} 1/u_k(x)$,
where the actual $u_k(x)$ can be assumed from the picture at the required position along the structure and the two outer bricks have been excluded. This method to compute $W_p$ is to be preferred to that based on eq. \eqref{ww_eq} as with the former it is possible to account for the effective data extracted from the snapshots and have an estimate of the strain energy as close as possible to the actual value. In this way, the strain energy turned out to be $W_p=0.340$ J.
The kinetic energy is
$
K_p=\mu \int_0^L v_b^2 \xi^2/L^2 d\xi,
$
where $\mu$ is the mass density per unit axis length that in our case is $\mu=m_b/h=0.345$ kg/m. After integration, $K_p=\mu v_b^2 L/3$. As $L=125$ mm, the equality $W_p=K_p$ yields $v_b \approx 4.86$ m/s.

\section{Concluding remarks}

The progressive instability behaviour of compressed dry-stone prismatic rectangular pillars made up of PMMA bricks of various heights and
loaded with eccentric loads was assessed experimentally with a photoelastic apparatus.
The experimental results were compared with the predictions obtained by the no-tension material model for which the columns are assumed linearly elastic in compression, but are incapable, due to the absence of the mortar, to withstand any tensile stress. While recalling the main closed-form results of the theory (i.e. load-lateral displacement curve, deformed shape of the axis of the beam-column, location of the interface between cracked and undamaged portions of the pillar), new analytical solutions for the position of the neutral axis in a generic cross section of the damaged part of the pillar and for the axial displacement of the structure were attained. With the latter, the validity of the Clapeyron's theorem was verified.

%The main conclusion of this work is that the no-tension material model is able to capture well the observed behaviour of the tested pillars for both low and high eccentricity loadings:
%while it slightly overestimates the global load-bearing capacity, it provides an accurate prediction of the deformed shape of the pillar and of the position of neutral axes in the damaged part of the structure.

Comparing the experimental outcomes with the theory, we conclude that the no-tension material model captures very well the observed behaviour of the tested pillars for both low and high eccentricity loadings.
In fact, while it slightly overestimates the global load-bearing capacity, it provides an
accurate prediction of the deformed shape of the pillar and of the position of neutral axes
in the damaged part of the structure.

%The collapse of a pillar subjected to a high-eccentric load was recorded with a high-speed camera and the evolution of a three-hinge failure mechanism documented.

Eventually, the collapse of the pillars subjected to a high-eccentric load was studied analytically and compared with the experiments.
The snapshots taken by a high-speed camera clearly show the evolution of the three-hinge failure mechanisms. The speeds of the central brick computed by using two energetic approaches still based on the no-tension model were in good agreement with the value determined experimentally.

\vspace{5 mm}

{\bf Acknowledgements.}
Support from the EU FP7 project ERC-AdG-340561-Insta\-bilities is gratefully acknowledged.

{\small

%\vspace{10mm}

%\setcounter{equation}{0}
%\renewcommand{\theequation}{{A}.\arabic{equation}}
%\begin{center}

%{\bf Appendix A - Useful derivatives of the invariants listed in Sect. \ref{sec_cost_eq}.}\\

%\end{center}

%\beq
%\derp{I_4}{\bF}=0, \ \ \ \
%\derp{I_5}{\bF}=2\bD\otimes\bD^0,\ \ \ \
%\derp{I_6}{\bF}=2(\bD\otimes\bC \bD^0+\bB \bD \otimes\bD^0),
%\eeq
%\beq
%\derp{I_4}{\bD^0}=2 \bD^0, \ \ \ \
%\derp{I_5}{\bD^0}=2\bC \bD^0,
%\ \ \ \ \derp{I_6}{\bD^0}=2 \bC^2 \bD^0,
%\eeq
%\beq
%\left(\derp{I_5}{\bF \partial \bD^0}\right)_{iJK}=2\delta_{JK}F_{iR}D^0_{R}+2 D^0_J F_{iK},
%\eeq
%\beq
%\left(\derp{I_6}{\bF \partial \bD^0}\right)_{iJK}=2 (F_{iK} C_{JR} D^0_{R}+C_{JK}F_{iR}D^0_{R}+F_{iR} C_{RK} D^0_{J}+
%\delta_{JK}F_{iR}C_{RS} D^0_{S}),
%\eeq
%\beq
%\left(\derp{\bB}{\bF}\right)_{ijkL}=\delta_{ik} F_{jL}+F_{iL} \delta_{jk}.
%\eeq

\end{document}